\documentclass[CRPHYS,Unicode,manuscript]{cedram}

\newcommand {\gtau} {$g^{(2)}(\tau)$ }

\newcommand {\el} {\mathrm{el}}
\newcommand {\bin} {\mathrm{bin}}
\newcommand {\coh} {\mathrm{c}}
\newcommand {\obs} {\mathrm{obs}}

\newcommand{\Lagrange}{Universit{\'e} C{\^o}te d'Azur, Observatoire de la C{\^o}te d'Azur, CNRS, Laboratoire Lagrange, France}
\newcommand{\INPHYNI}{Universit{\'e} C{\^o}te d'Azur, CNRS, Institut de Physique de Nice, France}

\title{Stellar intensity interferometry in the photon-counting regime}

\alttitle{Interférométrie d'intensité stellaire en comptage de photon}

\author{\firstname{William} \lastname{Guerin}\CDRorcid{0000-0001-8194-8351}\IsCorresp}
\address{\INPHYNI}
\email[W. Guerin]{william.guerin@univ-cotedazur.fr}
\author{\firstname{Mathilde} \lastname{Hugbart}\CDRorcid{0000-0002-5018-0236}}
\addressSameAs{1}{\INPHYNI}
\author{\firstname{Sarah} \lastname{Tolila}}
\addressSameAs{1}{\INPHYNI}
\author{\firstname{Nolan} \lastname{Matthews}
\CDRorcid{0000-0002-3687-4661}}
\addressSameAs{1}{\INPHYNI}
\address{Space Dynamics Laboratory, Utah State University, Logan, UT, 84341, USA}
\author{\firstname{Olivier} \lastname{Lai}\CDRorcid{0000-0001-5656-7346}}
\address{\Lagrange}
\author{\firstname{Jean-Pierre} \lastname{Rivet}\CDRorcid{0000-0002-0289-5851}}
\addressSameAs{2}{\Lagrange}
\author{\firstname{Guillaume} \lastname{Labeyrie}}
\addressSameAs{1}{\INPHYNI}
\author{\firstname{Robin} \lastname{Kaiser}\CDRorcid{0000-0001-5194-3680}}
\addressSameAs{1}{\INPHYNI}
\thanks{We acknowledge the financial support of the French National Research Agency (project I2C, ANR-20-CE31-0003). Part of this work was performed in the framework
of the European project IC4Stars, ERC Advanced Grant No. 101140677.}


\begin{abstract}
Stellar intensity interferometry consists in measuring the correlation of the light intensity fluctuations at two telescopes observing the same star. The amplitude of the correlation is directly related to the luminosity distribution of the star, which would be unresolved by a single telescope. This technique is based on the well-known Hanbury Brown and Twiss effect. After its discovery in the 1950s, it was used in astronomy until the 1970s, and then replaced by direct (``amplitude'') interferometry, which is much more sensitive, but also much more demanding. However, in recent years, intensity interferometry has undergone a revival. In this article, we present a summary of the state-of-the-art, and we discuss in detail the signal-to-noise ratio of intensity interferometry in the framework of photon-counting detection.
\end{abstract}

\begin{altabstract}
L'interférométrie d'intensité stellaire consiste à mesurer la corrélation entre les fluctuations d'intensité de la lumière captée par deux télescopes observant la même étoile. L'amplitude de la corrélation est directement liée à la taille de l'étoile, qui ne serait pas résolue par un seul télescope. Cette technique est basée sur le célèbre effet Hanbury Brown et Twiss. Après sa découverte dans les années 1950, elle a été utilisée en astronomie jusque dans les années 1970, puis remplacée par l'interférométrie directe (ou d'amplitude), qui est beaucoup plus sensible, mais aussi beaucoup plus exigeante techniquement. Cependant, depuis quelques années, l'interférométrie d'intensité connait un renouveau. Dans cet article, nous présentons un résumé de l'état de l'art et nous discutons en détail du rapport signal à bruit de l'interférométrie d'intensité dans le cadre de détecteurs en comptage de photons.
\end{altabstract}

\begin{document}

\maketitle

\section{Short review on stellar intensity interferometry}


\subsection{Historical work by Hanbury Brown and Twiss}

The history of intensity interferometry and photon correlation is closely related to the development of modern quantum optics.
After Einstein introduced the concept of light quanta, the way to reconcile this corpuscular picture with interference phenomena was to consider that photons always interfere with themselves. Until the 1950s, the notion of multiparticle interference was unknown to most physicists.
This view was deeply transformed thanks to the pioneering work of Hanbury Brown and Twiss (HBT)\cite{HBT:1956a,HBT:1956}, whose experiments challenged traditional ideas. Their results on intensity correlations were finally recognized as key milestones towards our understanding of the quantum nature of light, and triggered the development of the quantum theory of optical coherence \cite{Glauber:1963a, Glauber:1963b, Glauber:2006}.

The first demonstration of intensity correlation in the optical domain occurred with a laboratory demonstration \cite{HBT:1956a}. Shortly after, they applied this technique to starlight \cite{HBT:1956}, successfully measuring the angular size of the star Sirius. To this end, light was detected at two telescopes, as shown in the top-left picture of Fig.\,\ref{fig.HBT_HESS_VERITAS}, and then correlated as follows:
\begin{equation}
g^{(2)}(\tau,r) =  \frac{\langle I_1(t) I_2(r,t + \tau) \rangle} {\langle I_1 \rangle \langle I_2 \rangle}\,, \label{eq:g2}
\end{equation}
with $I_1$ and $I_2$ the intensity collected by the two telescopes, $\tau$ the time lag between them, $r$ the projected baseline between the telescopes, and with $\langle \,\cdot\, \rangle$ corresponding to the averaging over time $t$. 
For chaotic light, a classical explanation can relate this intensity correlation function to the electric field correlation function, 
\begin{equation}
g^{(1)}(\tau,r) =  \frac{\langle E_1^\star(t) E_2(r,t + \tau) \rangle} {\langle \sqrt{I_1 I_2} \rangle}\,. \label{eq:g1}
\end{equation}
The so-called Siegert relation\,\cite{Siegert:1943,Lassegues:2022} states:
\begin{equation}
g^{(2)}(\tau,r) = 1+|g^{(1)}(\tau,r) |^2    \label{eq:Siegert}
\end{equation}
with
\begin{equation}
g^{(1)}(\tau,r) = V(r)g^{(1)}(\tau), \label{eq:g1_Visibility}
\end{equation}
where $V(r)$ is the ``visibility'', i.e., the contrast of the interference fringes if the light from the two telescopes were combined. The first-order time correlation is directly given by the Fourier transform of the light power spectrum (Wiener-Khinchin theorem), while the visibility as a function of the spatial separation $r$ is the Fourier transform of the luminosity distribution of the source (van Cittert-Zernike theorem) \cite{MandelWolf}. For example, for a uniform disk distribution of angular diameter $\theta$,
\begin{equation}\label{eq.Airy}
|V(r)|^2 = \left[ \frac{2 J_1(\pi r\theta/\lambda)}{\pi r\theta/\lambda} \right]^2\,,
\end{equation}
where $J_1$ is the Bessel function of the first kind and $\lambda$ is the observation wavelength. The measurement of $g^{(2)}(\tau,r)$, therefore, provides spatial information on a source that would be unresolved with a single telescope, and without the need for direct interference.

However, these first results obtained by HBT were met with great skepticism (see \cite{Silva:2012} for a historical review of the controversy). A notable feature of the early HBT experiments was that they were performed with continuous wave detection. In this regime, photons were not required to explain the observed results, as the classical wave theory of light is sufficient. The instantaneous intensity $I(t)$ could be interpreted as the probability of detecting light at a given moment, without invoking the concept of discrete photons. Nevertheless, when intensity interferometry was considered in the photon-counting regime, the classical framework could no longer fully account for the results, and a quantum description became necessary.

In 1956, Purcell provided a theoretical insight: photon bunching, observed in intensity correlation measurements with $g^{(2)}(r=0,\,\tau=0) > 1$, was a natural consequence of the Bose-Einstein statistics obeyed by photons as bosons \cite{Purcell:1956}. HBT further developed this interpretation \cite{HBT:1957a}, as did Kahn \cite{Kahn:1958}, and the effect was experimentally confirmed in a photon-counting experiment by HBT themselves \cite{HBT:1957} and by Pound \& Rebka \cite{Rebka:1957}. Slightly later, Fano introduced another interpretation, framing the phenomenon in terms of two-photon interference \cite{Fano:1961}. Together, these contributions deepened the understanding of photon statistics and light’s quantum nature, and intensity correlation is now a tool that is widely used to probe the quantum nature of light.

The use of spatial intensity correlations was then widely applied by HBT in the 1960s and 1970s. Since the amplitude of the bunching peak $g^{(2)}(r, \tau=0)$, obtained by correlating the light detected by two telescopes separated by a projected baseline $r$, is directly proportional to the squared modulus of the visibility (Eqs.\,\ref{eq:Siegert} and \ref{eq:g1_Visibility}), one has access to the star's diameter. This approach allowed the measurement of the angular diameter of 32 stars using the Narrabri Stellar Intensity Interferometer in Australia \cite{HBT:1974}. 

Despite its success, the technique was ultimately replaced by amplitude interferometry, pioneered by A. Labeyrie, who obtained first fringes on Vega in 1975 at Calern (France) \cite{Labeyrie:1975}. Unlike stellar amplitude interferometry, which involves directly combining light from multiple telescopes to produce interference fringes and demands precise control of path-length differences at the scale of an optical wavelength, stellar intensity interferometry (SII) is phase-insensitive and simpler to implement. However, this simplicity comes at the cost of significantly lower sensitivity (i.e. signal-to-noise ratio) compared to amplitude interferometry, which has, to date, restricted its application to the observation of very bright stars \cite{HBT:1974}.

\subsection{Modern revival of intensity interferometry}

Thanks to advancements in photonic components, efficient single-photon counting detectors, fast electronics, and digital correlators, there is currently significant interest from various research groups in reviving stellar intensity interferometry (SII) using modern photonic technologies. One promising approach to implementing stellar intensity interferometry involves Cherenkov telescope arrays, principally designed for ground-based gamma-ray astronomy, such as CTAO (see, e.g., \cite{Dravins:2013}, and references therein). 

These arrays offer two key advantages. First, their large collectors, with diameters of more than 10 meters, enable the capture of a significant number of photons, improving the signal-to-noise ratio and allowing, in particular, access to fainter stars. Secondly, their spatial arrangement enables correlation measurements across a wide range of baselines with a two-dimensional distribution. While intensity interferometers can only measure the squared modulus of the electric field's correlation function — thereby losing phase information and making direct star image reconstruction impossible\footnote{Accessing the Fourier phase is in principle feasible by measuring third order correlations \cite{Wentz:2014, Nunez:2015}, but this has not yet been demonstrated due to a very low signal-to-noise ratio.} — the use of an interferometer with varying baselines in two-dimensional space can impose significant constraints on the source image\,\cite{Dravins:2013}.

\begin{figure}[t]
    \centering
    \includegraphics[width=0.9\linewidth]{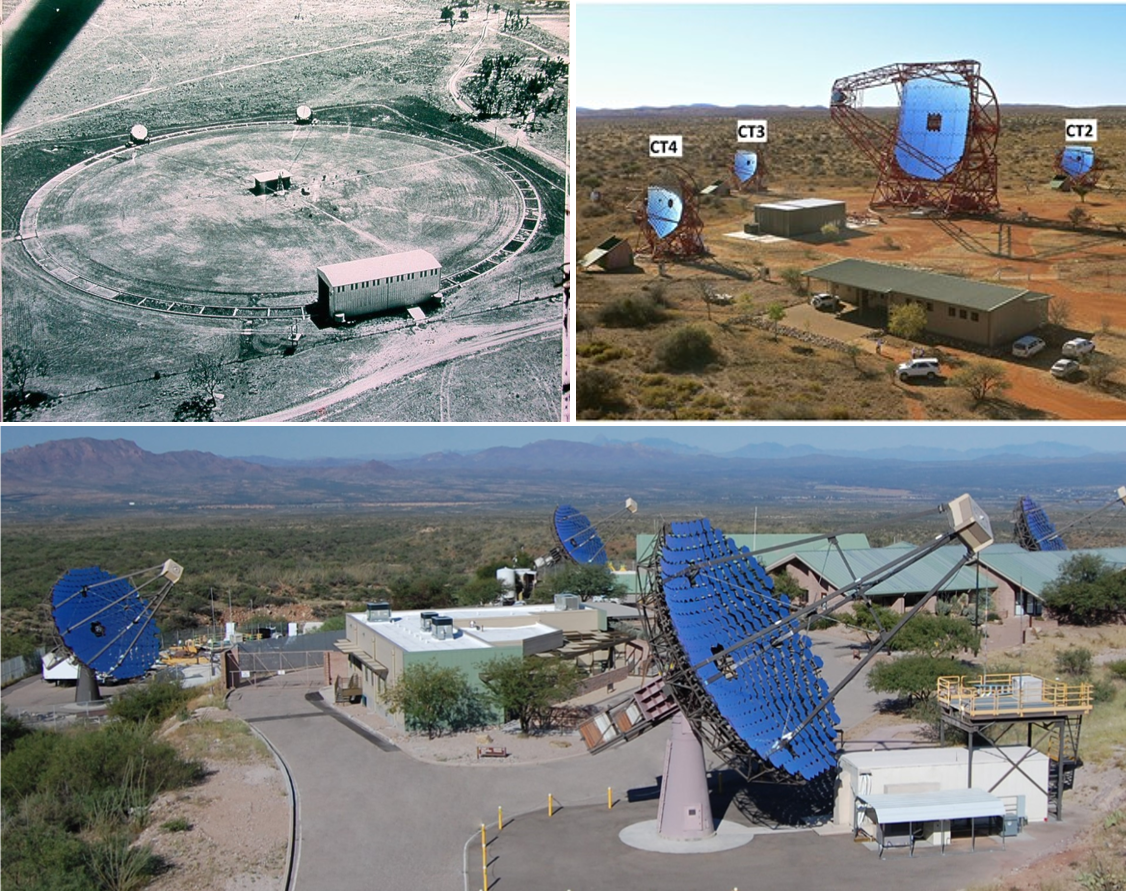}
    \caption{Pictures of different facilities used for intensity interferometry with large light collectors. Top-left: the Narrabri observatory, with two telescopes set up on railroad tracks to vary the baseline size and orientation \cite{HBT:1968}. Top-right: the H.E.S.S. site in Namibia with Cherenkov telescopes 3 and 4 specifically used for intensity interferometry\,\cite{Zmija:2023}. Bottom: VERITAS array in Arizona with 4 Cherenkov telescopes\,\cite{Abeysekara:2020}.}
    \label{fig.HBT_HESS_VERITAS}
\end{figure}

Recently, successful measurements have been done using Cherenkov telescopes, with currently three main observatories operating with at least two telescopes each: H.E.S.S., MAGIC, and VERITAS. The H.E.S.S. array in Namibia, consisting of 12 m diameter telescopes shown in the top-right picture in Fig.\,\ref{fig.HBT_HESS_VERITAS}, measured the spatial coherence curves of two stars with two telescopes, achieving angular diameter precision at the 10$\%$ scale\,\cite{Zmija:2023}, and demonstrated dual-band (two colours) interferometry \cite{Vogel:2025}. The MAGIC system is located at the Roque de los Muchachos Observatory in La Palma, Spain, and features two telescopes with 17 m diameter parabolic reflectors. It enabled the first measurement of 13 stellar diameters in the 400–440\,nm band and a total of 22 stellar diameters \cite{Acciari:2020,MAGIC:2024}. Its sensitivity allowed for relative errors of a few percent, which provides valuable astrophysical parameters beyond stellar diameters, such as the oblateness of certain stars, yielding constraints on their rotational speeds, a key factor influencing their structure and evolution \cite{MAGIC:2024}. Finally, VERITAS, a four-telescope array in Arizona with 12 m diameter reflectors, with projected baselines between 35 and 170 m as shown in the bottom of Fig.\,\ref{fig.HBT_HESS_VERITAS}, measured the angular diameters of several stars, and for some of them for the first time at visible wavelengths (416 nm)\,\cite{Abeysekara:2020, Acharyya:2024}. For example, their uncertainty of less than 4\,$\%$ allowed them to show that the mean uniform disk angular diameter of $\beta$\,UMa is smaller than those observed at longer wavelengths, consistent with the expected stronger limb darkening at shorter wavelengths.

A limitation of Cherenkov telescopes is their poor imaging quality, with typical angular point-spread functions (PSF) of $\sim 0.1^o$, corresponding to focal spot sizes of several mm or even cm. This forces the use of large-area detectors (such as photomultiplier tubes) and sets a sensitivity limit due to the sky background \cite{Rou:2013}. 

An alternative approach involves the use of optical telescopes of imaging quality \cite{Rivet:2018}. This method offers several advantages, including a much smaller PSF, which enables efficient coupling of light into optical fibers. This also facilitates the use of high quantum efficiency detectors with high timing resolution, along with narrow filters to enhance the contrast of the bunching peak, isolate emission lines, and reduce sky background. However, the number of photons collected is lower compared to Cherenkov telescopes, as optical telescopes typically have smaller collecting area. Consequently, the sensitivity of this technique must be significantly improved to achieve competitive performance.


Several groups have adopted this approach. Besides our group in Nice (see the next section), this track has been followed by the Asiago Observatory (Italy). In 2021, using the 1.22-m Galileo telescope and the 1.82-m Copernicus telescope, separated by nearly 4 km, they successfully demonstrated the temporal correlation of Vega at zero baseline, along with measurements of correlation on a projected baseline of approximately 2 km (but showing no bunching peak as the star is fully resolved) \cite{Zampieri:2021}. Another experiment was conducted in 2022 at the Southern Connecticut Stellar Interferometer using three portable 0.6-m Dobsonian telescopes equipped with single-photon avalanche diode detectors. Observations with two telescopes detected a correlation peak when observing unresolved stars, and with results consistent with partial or no detection when using a baseline sufficient to resolve the stars \cite{Horch:2022}. Finally, an experiment was carried out at the 1.04-m Omicron telescope of the C2PU (Centre Pédagogique Planète Univers) at Calern Observatory (France) by a team from Erlangen University in 2023. This experiment achieved the successful observation of photon bunching for three bright stars, utilizing hybrid photon detectors designed for single-photon detection with a large detection area and high timing resolution \cite{Leopold:2024}. More recently, they also observed spatial correlation between the two C2PU telescopes.

Lastly, while one of the prime goals of the use of optical telescopes is to reach faint magnitudes (up to magnitude 8 for current experiments, higher in the future) and sub-milliarcesond resolution, requiring baselines of up to a few kilometers (see, in particular, the project of a Swiss consortium \cite{Walter:2023}), a complementary strategy involves the use of large arrays of small telescopes \cite{Mozdzen:2025}. Although the smaller aperture limits the signal-to-noise ratio and restricts observations to very bright stars, the ability to construct an array with a large number of telescopes and numerous short baselines could enable precise image reconstruction of bright giant stars \cite{Nunez:2012a,Nunez:2012b,Dravins:2015b}. Moreover, a dedicated array could be used for long observing programs to reach fainter magnitudes, as the signal from individual observations can be combined post-facto. Let us finally note that this idea of using many small apertures has also been proposed for Cherenkov telescopes by exploiting individual facets as independent sub-apertures and applying aperture synthesis techniques using pairwise and triple intensity correlations \cite{Gori:2021}.

\subsection{Results of the Nice group}

\begin{figure}[t]
    \centering
    \includegraphics{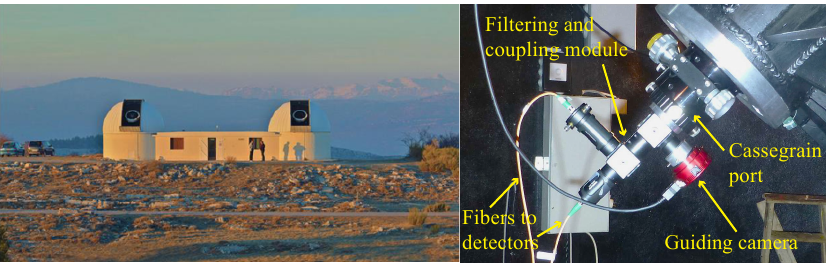}
    \caption{Intensity interferometry performed by the I2C consortium at Calern Observatory (France). Left: The two telescopes of C2PU, separated by 15\,m, used for spatial intensity interferometry. Right: The coupling device connected to the telescope’s Cassegrain port. It spectrally filters the incoming light, separates the two polarization channels, and injects the light into multimode fibers that feed photon detectors \cite{deAlmeida:2022}.}
    \label{fig.C2PU}
\end{figure}

Our group, a collaboration of physicists and astronomers within the Université Côte d’Azur network, began its work on SII in 2017, building on earlier research using intensity correlations to study light scattered by hot atomic vapors \cite{Dussaux:2016}. Since then, we have achieved several notable milestones. Thanks to the available facilities at Calern Observatory, we first observed the bunching peak in stellar light using a single telescope \cite{Guerin:2017} and later performed SII with the two C2PU telescopes (15\,m baseline) shown on the left of Fig.\,\ref{fig.C2PU} \cite{Guerin:2018}. These observations marked the first successful intensity interferometry measurements of stars\footnote{With a single telescope, photon bunching was observed with Sun light slightly earlier by Tan and coworkers \cite{Tan:2014}.} since the Narrabri Observatory era \cite{HBT:1974} and were the first to use photon-counting detection.

By employing narrow filters centered on emission lines, we isolated fine spectral features, giving insights into stellar physics. For instance, we conducted SII measurements on the H$\alpha$ line at $\lambda = 656.3$\,nm for stars such as P Cygni \cite{Rivet:2020,deAlmeida:2022}, Rigel \cite{deAlmeida:2022}, and $\gamma$ Cas \cite{Matthews:2023}, confirming and refining earlier results from amplitude interferometry. When combined with spectroscopic observations and radiative transfer modeling, this can also provide a precise distance estimate, as done, for example, for P Cygni and Rigel with a slight adjustment of their distance \cite{deAlmeida:2022}, paving the way towards an independent distance indicator for extragalactic sources \cite{Kudritzki:1999}.

Thanks to its compact design, as shown on the right picture of Fig.\,\ref{fig.C2PU}, our SII apparatus is portable and adaptable to various large telescopes. Demonstrating this versatility, we carried out observation campaigns with a range of instruments: a 4.1 m telescope at SOAR in Chile \cite{Guerin:2021}, the MéO 1.5 m telescope and a portable 1 m telescope at Calern Observatory \cite{Matthews:2023}, and two missions at the VLTI in Chile, using Auxiliary Telescopes (ATs) with baselines of 49\,m and later three ATs for more advanced measurements \cite{Matthews:2022}.

One long-term goal is to develop an instrument capable of performing interferometric measurements across visible wavelengths (B, V, R, I bands) with sensitivity to stars brighter than magnitude 8 when deployed on major facilities. Achieving this goal requires a significant enhancement of the sensitivity of the current instrument, for instance, through wavelength multiplexing and the use of even better detectors (see Section \ref{sec.future}). A critical aspect in evaluating the future potential is the determination of the signal-to-noise ratio. 

\section{The signal-to-noise ratio with single-photon detectors}

After this overview of the state of the art, in this second part, we discuss in detail the signal-to-noise ratio (SNR). We first recall HBT's well-known formula and its validity conditions, and then how to adapt it and derive it in the photon-counting regime. We show numerical and experimental data that validate our SNR equation. We also discuss experimental artifacts to be avoided to reach the theoretical SNR and some limitations that are hard to avoid. Finally, we discuss the on-going work to significantly enhance the sensitivity by implementing wavelength multiplexing with high-time resolution detectors.

\subsection{HBT's formula}\label{sec.SNR_HBT}

The SNR has been computed by HBT in a way adapted to their method for measuring the correlations, i.e. with analogue electronics. They give a rather complicated expression in Ref.\,\cite{HBT:1958} (Eq.\,4.1), which takes into account many factors, for instance, the partial loss of correlation due to the finite aperture of individual telescopes, the exact frequency response of the electronics, the extra noise from the electronics, etc. This complete expression is hardly used and is often simplified and quoted as:
\begin{equation}\label{eq.SNR_Deltaf}
S\!N\!R = \alpha  N_\mathrm{ph}(\lambda) A |V(r)|^2 \sqrt{\frac{T_\mathrm{obs} \Delta f}{2}} \,.
\end{equation}
Here, the different factors are:
\begin{itemize}
\item $N_\mathrm{ph}(\lambda)$ the photonic spectral flux from the star at wavelength $\lambda$, i.e. a number of photons per second, per squared meter and per Hertz of optical bandwidth (unit: ph\,s$^{-1}$\,m$^{-2}$\,Hz$^{-1}$).
\item $A$ the collecting area of each telescope, supposed to be equal. If they are not equal, the geometric mean should be used.
\item $\alpha$ the total throughput of the detection system, which includes all losses in the telescope and the detection efficiency of the detectors. If $N_\mathrm{ph}(\lambda)$ is given before the atmosphere, it should also include the attenuation of the latter.
\item $|V(r)|^2$ the squared visibility, which is actually what we want to measure. It depends on the source, the projected baseline $r$ of the telescopes, and the wavelength (Eq.\,\ref{eq:g1_Visibility}).
\item $T_\mathrm{obs}$ the observation time.
\item $\Delta f$ the electronic bandwidth (-3dB).
\end{itemize}
This simplified formula relies on a number of assumptions:
\begin{itemize}
\item the incident light is unpolarized;
\item spurious light and dark current (dark counts for single-photon detectors) are negligible;
\item spurious correlations are negligible;
\item the electronics act as a low-pass filter of infinite order and do not introduce any other noise (i.e., the only noise is the shot noise);
\item the coherence time is not resolved, i.e. $\tau_\coh \ll 1/\Delta f$, or equivalently, $\Delta\nu \gg \Delta f$.
\end{itemize}
This last assumption explains why the SNR is independent of the optical bandwidth $\Delta\nu$. Increasing $\Delta\nu$ also increases the number of detected photons but decreases the temporal coherence $\tau_\coh =1/\Delta\nu$. Since the measured $g^{(2)}(\tau)$ function is the convolution of the true $g^{(2)}(\tau)$ before instrumental effect, of width $\tau_\coh$, with the electronic response of width $1/\Delta f$, the contrast of the bunching peak is reduced to $\sim \Delta f /\Delta \nu \ll 1$. The two effects compensate exactly. In practice, spectral filtering is necessary to fulfill the above assumptions. Indeed, a too large spectrum may reduce the coherence time so much that the amplitude of the convolved bunching peak may become comparable to spurious correlations, and also the detectors may saturate. Moreover, the visibility depends on the wavelength (Eq.,\ref{eq.Airy}), so a broad spectrum (large $\Delta\lambda$) makes the interpretation of the results more complex. On the other hand, too narrow filtering may reduce the photon rate from the source so much that the dark count or spurious light may become nonnegligible. Typically, spectral filtering between one and a few nm is used. Another remark is that polarizing the light also does not change the final SNR: it increases the bunching peak by a factor two, but decreases $\alpha$ by the same factor, corresponding to the transmission through the polarizer\footnote{Processing independently the two polarization channels leads to a $\sqrt{2}$ increase of the SNR at the expense of doubling the number of detectors \cite{deAlmeida:2022,Matthews:2022,Matthews:2023}.}. Finally, this equation is adapted to the case of measuring correlation between two telescopes, each of collecting area $A$. If, for calibration purpose, one measures the so-called zero-baseline correlation, i.e. the $g^{(2)}(\tau, r=0)$ at a single telescope by splitting the light onto two detectors (like in our first experiment \cite{Guerin:2017}), then the SNR of that measurement is divided by two.

The factor $\alpha$ is often not precisely known, as it depends on many factors such as the transmission through the atmosphere and through the complete optical system (telescope, coupling optics, optical fibers, etc.). $N_\mathrm{ph}$ may also be unknown, for example with an artificial source in the lab. It is thus useful to write the SNR as a function of the detected photon count rate. If one supposes that the optical filter has a square transmission spectrum of width $\Delta\nu$ and that the source spectrum is constant over the passband of the filter, the detected flux (count/s or cps) per telescope is
\begin{equation}
F = \alpha  N_\mathrm{ph}(\lambda) A \Delta\nu \,.
\end{equation}
This yields
\begin{equation}\label{eq.SNR_Deltaf2}
S\!N\!R = \frac{F}{\Delta\nu} |V(r)|^2 \sqrt{\frac{T_\mathrm{obs} \Delta f}{2}} \,.
\end{equation}
In the more general case of two telescopes and fluctuating count rates (due to seeing and scintillation), the SNR will be proportional to $<\sqrt{F_1(t) F_2(t)}>$ (average over time), which can be slightly smaller than $\sqrt{<F_1(t)> <F_2(t)>}$ in case of strong uncorrelated fluctuations.

\subsection{Adaptation to the photon-counting regime}

In recent years, this formula has been often used by replacing $\Delta f$ with an ``electronic timing resolution'' $\tau_\el$, a terminology more appropriate in the photon-counting regime. Most of the time, it has been said that $\Delta f \sim 1/\tau_\el$, without precisely defining $\tau_\el$ (for example: rms or FWHM?) \cite{Guerin:2017, Karl:2022} or with a precise definition, e.g. the time bin, but without questioning the numerical prefactor \cite{Zampieri:2021}. In this section, our goal is to clarify this point.

In a very recent paper\footnote{We are not aware of any previous publication giving the SNR with these exact numerical factors.}, Dalal \textit{et al}. derive the analytical formula \cite{Dalal:2024}:
\begin{equation}\label{eq.SNR_perimeter}
S\!N\!R = \frac{1}{2} \alpha N_\mathrm{ph}(\lambda) A |V(r)|^2 \sqrt{\frac{T_\mathrm{obs}}{2\sqrt{\pi}\tau_\mathrm{el}}} =  \frac{|V(r)|^2}{2} \frac{F}{\Delta\nu} \sqrt{\frac{T_\mathrm{obs}}{2\sqrt{\pi}\tau_\mathrm{el}}} \; ,
\end{equation}
with $\tau_\el$ the rms width of the timing resolution of the correlation (i.e., $\sqrt{2}$ times the timing resolution of a single detector), assuming a Gaussian jitter distribution function for the detectors. Compared to HBT's formula\footnote{It is easy to compute the Bode plot corresponding to a Gaussian jitter, and one finds that the -3\,dB bandwidth is $\Delta f = \sqrt{\ln(2)}/(2\pi\tau_\el)$, a very slightly lower value than $\Delta f = 1/(4\sqrt{\pi} \tau_\el)$. This is because Eq.\,(\ref{eq.SNR_Deltaf}) is simplified (only valid for an infinite-order filter, see. \cite{HBT:1958}).}, the electronic bandwidth $\Delta f$ has been replaced by $(4\sqrt{\pi} \tau_\el)^{-1}$. We have highlighted the factor $1/2$ due to the fact that light is unpolarized, it should be removed if light is polarized.

The derivation by the authors is rather theoretical and does not relate to a specific measurement protocol. In the following, we recover this result by supposing that the bunching peak amplitude is determined by a Gaussian fit with only one free parameter. We will also check this result numerically. Furthermore, we investigate the effect of having more fitting parameters or using a numerical integration instead of a fit. Finally, we will compare this formula with experimental data taken on an artificial star in the laboratory.

\subsubsection{Alternative derivations of the SNR}

Let us consider an HBT setup with two detectors in the photon-counting regime, each connected to a time-to-digital convertor (TDC, or time tagger) that records the arrival time of each detected photon (see Fig.\,\ref{fig.spurious_correlations}(a) for an illustration). The unnormalized $G^{(2)}(\tau)$ is the histogram of all delays $\tau$ between counts on the two detectors. In the absence of any correlation (i.e. far from the bunching peak), on average, the accumulated number of coincidences in each time bin of the histogram is 
\begin{equation}\label{eq.Nc}
N_\mathrm{c} = F_1 F_2 T_\obs t_\bin \, ,
\end{equation}
where $F_1, F_2$ are the average count rates on each detector, $T_\obs$ is the integration time and $t_\bin$ is the time bin of the histogram. The noise is $\sqrt{N_\mathrm{c}}$ and, after normalization by $N_\mathrm{c}$, the noise on the \gtau function has an rms amplitude:
\begin{equation}\label{eq.noise}
\sigma_\mathrm{noise} = \frac{1}{F \sqrt{T_\obs t_\bin}} \, ,
\end{equation}
where $F$ is the geometric average of the count rates.

On the other hand, the signal is the amplitude of the bunching peak that we want to measure. To simplify the notation, let us suppose that the light has been polarized and that the spatial coherence is maximal (unresolved star). This amounts to removing the factor $|V(r)|^2/2$, which can be added again at the end. Let us also suppose an optical filtering $s(\nu)$ with a square spectrum of width $\Delta\nu$. The theoretical \gtau function is then (Eq.\,\ref{eq:Siegert} and Wiener-Khinchin theorem): 
\begin{equation}\label{eq.sinc}
g^{(2)}(\tau) = 1+ |g^{(1)}(\tau)|^2 = 1+ \left(\frac{\sin(\pi \Delta\nu \tau)}{\pi \Delta\nu \tau}\right)^2 \, .
\end{equation}
However, this theoretical function will be convolved by the electronic response of the detection chain. In this process, the integral of $g^{(2)}(\tau) - 1$ is conserved. This integral of the bunching peak is the coherence time, which is only related to the optical filter:
\begin{equation}\label{eq.tau_c}
\tau_\coh = \int |g^{(1)}(\tau)|^2 d\tau = \int |s(\nu)|^2 d\nu = \frac{1}{\Delta\nu} \,.
\end{equation}

Now we will start making a few hypotheses to simplify the problem. First, the response function of the detector is Gaussian, i.e. each detected photon suffers from a time uncertainty (jitter) with a Gaussian distribution. For the correlation, this also gives a Gaussian convolution of rms width $\tau_\el$, which is $\sqrt{2}$ larger than the rms jitter of individual detectors. Second, we assume that $\tau_\el \gg \tau_\coh$, which has been the case for all experimental setups in astronomy so far (except on the Sun \cite{Tan:2014}). Then the initial shape of the \gtau function is lost in the convolution, only the integral remains. The measured bunching peak is thus a Gaussian of integral $\tau_\coh$ and width $\tau_\el$. Its amplitude (or contrast) is thus
\begin{equation}\label{eq.ampl}
C = \frac{\tau_\coh}{\sqrt{2\pi} \tau_\el} = \frac{1}{\sqrt{2\pi} \tau_\el \Delta\nu} \,.
\end{equation}
Also, since $C \ll 1$, the number of coincidences in the bunching peak is not significantly different from the ones at a larger time lag; therefore, we can consider that the amplitude of the noise is the same.

The problem is now the following: we have a Gaussian of amplitude $C\ll 1$, rms width $\tau_\el$, with an offset of 1, a time sampling $t_\bin$ and a Gaussian noise of rms amplitude $\sigma_\mathrm{noise}$. Remember that $C$ is given by Eq.\,(\ref{eq.ampl}) multiplied by the set-aside factor $|V|^2/2$, which is the quantity we are interested in. Therefore, the sought-after SNR is given by the statistical uncertainty $\sigma_C$ in the determination of $C$: $S\!N\!R=C/\sigma_C$ (or, equivalently, in the determination of the integral).
\\

\noindent\emph{Integral method --} Let us discuss a first method: the direct numerical integration of the data to determine the bunching-peak area. It has the advantage of not relying on any prior knowledge of the response function of the detectors. The drawback is that it does not lead to the best SNR. Indeed, the integral will give $\tau_\coh = 1/\Delta\nu$ on average with a statistical uncertainty that is directly the uncertainty $\sigma_\mathrm{int}$ of the integral of the noise (here 0), given the number of points in the distribution, $n=T_\mathrm{w}/t_\bin$, where $T_\mathrm{w}$ is the window of integration: $\sigma_\mathrm{int} = \sigma_\mathrm{noise}\sqrt{n}t_\bin$. This gives an SNR
\begin{equation}\label{eq.SNR_int}
S\!N\!R_\mathrm{int} = \frac{F}{\Delta\nu} \sqrt{\frac{T_\obs}{T_\mathrm{w}}} \, .
\end{equation}
As intuitively expected, it does not depend on $t_\bin$ and $\tau_\el$ any more, but only on $T_\mathrm{w}$. Since one needs $T_\mathrm{w}$ significantly larger than $\tau_\el$ to get the full integral, this method is not favorable. Note that if the time bin is the limiting factor of the whole detection chain, then the bunching peak is just one point and this formula applies with $T_\mathrm{w} = t_\bin$.
\\

\noindent\emph{Binning method --} Another approach to determine the height of the bunching peak is simply to look for the maximum value in the \gtau function, and take the rms noise in the uncorrelated part as the uncertainty. The result obviously depends on the binning and the question is thus to find the optimal binning $T_\bin$. The SNR is given by:
\begin{equation}
S\!N\!R_\mathrm{bin} = \frac{F}{\Delta\nu} \sqrt{\frac{T_\obs}{T_\bin}} \times \mathrm{erf}\left( \frac{T_\bin}{2\sqrt{2}\tau_\el} \right)\, .
\end{equation}
Compared to the previous equation, $T_\mathrm{w}$ has been replaced by $T_\bin$ and the extra erf factor accounts for the fact that the integral is not taken over the full \gtau function. Numerically, we find that this SNR is maximum for $T_\bin \approx 0.99 \times 2\sqrt{2} \tau_\el \approx 2.8 \tau_\el$, which gives 
\begin{equation}
S\!N\!R_\mathrm{bin} \approx \mathrm{erf}(1) \times \frac{F}{\Delta\nu} \sqrt{\frac{T_\obs}{2\sqrt{2}\tau_\el}} \approx 0.5 \times \frac{F}{\Delta\nu} \sqrt{\frac{T_\obs}{\tau_\el}} .
\end{equation}
This is slightly lower than the SNR of Eq.\,(\ref{eq.SNR_perimeter}), as $1/\sqrt{2\sqrt{\pi}} \approx 0.53$ (we recall that we have removed the factor $|V|^2/2$). Moreover, this method suffers from the same caveats as the one-parameter fit, as discussed in detail in Sec.\,\ref{sec.practical}.
\\

\noindent\emph{Fit method --} Let us now determine $C$ by a least-square fitting. The uncertainty is
\begin{equation}
\sigma_C = \sqrt{\frac{SS}{DF}\mathrm{cov}(j,j)} \;,
\end{equation}
where $SS$ is the sum of squared residuals (i.e. the sum of the squared difference between the data points and the fit), $DF$ is the number of degrees of freedom, equal to the number of data points minus the number of fitting parameters, and $\mathrm{cov}(j,j)$ is the diagonal element of the covariance matrix corresponding to the fitting parameter $j$. To go further analytically, we will now consider that there is only one fitting parameter, $C$. Then,
\begin{equation}
\mathrm{cov}(j,j)^{-1} = \sum_i \left( e^{-t_i^2/(2\tau_\el^2)} \right)^2 = \sum_i e^{-t_i^2/\tau_\el^2} \;,
\end{equation}
where the $t_i$'s are the data points. We also consider that the number of data points $n$ is large so that $DF \simeq n$. We then have $\frac{SS}{DF} = \sigma_\mathrm{noise}^2$, the variance of the noise. Therefore, 
\begin{equation}
\sigma_C = \frac{\sigma_\mathrm{noise}}{\sqrt{\sum_i e^{-t_i^2/\tau_\el^2}}} \;.
\end{equation}
In the hypothesis of a very good sampling, i.e. $t_\bin \ll \tau_\el$, we can approximate the sum by an integral:
\begin{equation}
\sum_i e^{-x_i^2/\tau_\el^2} \simeq \frac{1}{t_\bin} \int e^{-t_i^2/\tau_\el^2} dt = \sqrt{\pi} \frac{\tau_\el}{t_\bin}\;.
\end{equation}
This yields
\begin{equation}
\sigma_C = \frac{\sigma_\mathrm{noise}}{\pi^{1/4}} \sqrt{\frac{t_\bin}{\tau_\el}} \;,
\end{equation}
which finally gives, using Eqs.\,(\ref{eq.noise}) and (\ref{eq.ampl}), 
\begin{equation}\label{eq.SNR_fit}
S\!N\!R_\mathrm{fit} = \frac{F}{\Delta\nu} \sqrt{\frac{T_\obs}{2\sqrt{\pi}\tau_\el}} \, ,
\end{equation}
in agreement with Eq.\,(\ref{eq.SNR_perimeter}). We have thus recovered the result of \cite{Dalal:2024} by using the following hypothesis on the measurement protocol: only one fitting parameter (i.e., prior knowledge of $t_0$ and $\tau_\el$, the position and width of the bunching peak) and $t_\bin, \tau_\coh \ll \tau_\el$.

\subsubsection{Numerical investigation}

\begin{figure}[b]
\includegraphics{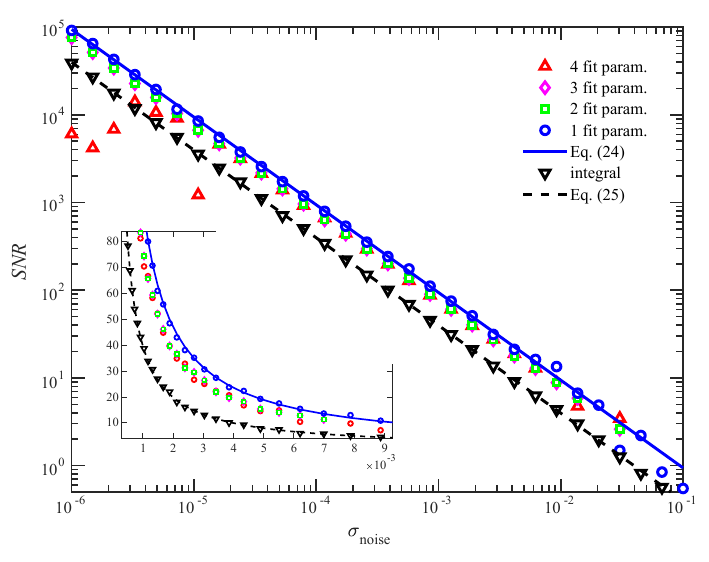}
\caption{Signal-to-noise ratio in the determination of the bunching peak amplitude, as a function of the rms amplitude of the noise, $\sigma_\mathrm{noise}$, in log-log scale. Inset: the same in linear scale for a restricted range. The fitting parameters are the amplitude $C$ (for all), the rms width $\tau_\el$ (for two or more fitting parameters), the center of the peak $t_0$ (three or four fitting parameters), and the offset $y_0$ (only with four fitting parameters). For the integral method, we have computed the integral over many realizations of the noise and plotted the ratio between the mean and the standard deviation.}
\label{fig.SNR_num}
\end{figure}

Let us check the previous results numerically. We consider the following function
\begin{equation} \label{eq.gaussian}
g(t_i) = y_0 + C \exp\left(-\frac{(t_i-t_0)^2}{2\tau_\el^2}\right) + X_i
\end{equation}
with $y_0 = 1$, $t_0=0$, $\tau_\el = 1$, $C=10^{-2}$ (corresponding to $\tau_\coh \simeq 0.025$), and $X_i$ a normally-distributed random variable of mean 0 and standard deviation $\sigma_\mathrm{noise}$. Moreover, we consider a sampling with $t_\bin = 0.02$ and a total time window $[-T_\mathrm{w}/2, \;\; T_\mathrm{w}/2]$ with $T_\mathrm{w} = 20$ ($10^3$ data points). We vary $\sigma_\mathrm{noise}$ for a fixed $t_\bin$. Therefore, we use Eqs.\,(\ref{eq.ampl}) and (\ref{eq.noise}) to change Eq.\,(\ref{eq.SNR_fit}) to:
\begin{equation}\label{eq.SNR_num}
S\!N\!R_\mathrm{fit} = \frac{C}{\sigma_C} = \frac{C}{\sigma_\mathrm{noise}} \sqrt{\frac{\sqrt{\pi}\tau_\el}{t_\bin}} \;.
\end{equation}
Note that for real data, $\sigma_\mathrm{noise}\propto 1/\sqrt{t_\bin}$, so that the result is well independent of $t_\bin$. Also, we recall that $C \propto 1/\tau_\el$, such that the SNR decreases with $\sqrt{\tau_\el}$.

Similarly, we can verify the result of the integral procedure, by verifying the Eq.\,(\ref{eq.SNR_int}), changed (using Eqs.\,\ref{eq.ampl} and \ref{eq.noise}) to
\begin{equation}\label{eq.SNR_num}
S\!N\!R_\mathrm{int} = \frac{C}{\sigma_\mathrm{noise}} \frac{\sqrt{2\pi}\tau_\el}{\sqrt{t_\bin T_\mathrm{w}}} \;.
\end{equation}

Figure \ref{fig.SNR_num} shows such comparisons, along with the numerical results with two, three and four fitting parameters. We can check that the analytical results are valid and that the one-parameter fitting procedure is the most efficient. Using four fitting parameters, i.e.$y_0$, $t_0$, $\tau_\el$ and $C$ (Eq.\,\ref{eq.gaussian}), is clearly bad: the SNR degrades at very high SNR and the fitting procedure is less robust at low SNR, leading to a few outlier points, for which the fit did not converge properly. Interestingly, using three ($C$, $\tau_\el$ and $t_0$) or only two ($C$ and $\tau_\el$) fitting parameters does not make any difference. However, it is significantly better to use only one fitting parameter ($C$). For example, for $\sigma_\mathrm{noise}\simeq 5\times10^{-3}$, the SNR is 15 with two or three fitting parameters and 20 with only one (and 8 for the integral method).

\subsection{Experimental investigation}

We now turn to the application of the previous results to real-world setups.

\subsubsection{Avoiding spurious correlations}

A major issue in experiments is to avoid any spurious correlations. When using continuous detectors and analog electronics, the main source of spurious correlation is electronic pick-up of radio frequency noise (see, for instance, ref. \cite{Kieda:2022}). One could naively think that using detectors in the photon-counting regime (such as single-photon avalanche diodes, SPADs) and digital electronics should prevent any spurious correlation. This is not the case, and we have identified four sources of spurious correlations.

\begin{figure}[t]
\includegraphics{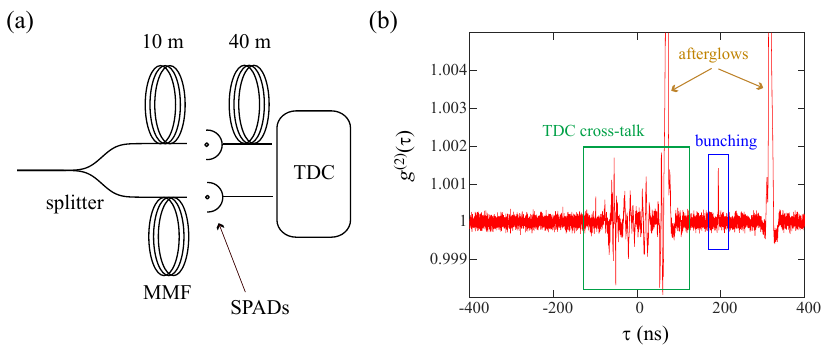}
\caption{(a) Scheme of the setup to measure \gtau without spurious correlations. At the entrance of the fibered splitter, we inject light coming from a white-light source, which has been subsequently injected in a single-mode fiber (to ensure maximum spatial coherence), spectrally filtered (with $\lambda=780$\,nm and $\Delta\lambda = 1$\,nm), and polarized (not shown here, see e.g. \cite{Guerin:2017}). The 10-m multimode fibers (MMF) between the splitter and the SPADs ensures that the afterglow peaks are well separated from the bunching peak, and the 40-m coaxial cable on one channel ensures that the bunching peak is far away from the TDC cross-talk. Finally, the two SPADs are separated by $\sim 3$\,m to avoid radiative cross-talk. (b) Corresponding \gtau measurement, with an integration time of 46\,h and count rates of $1.6\times 10^6$ counts per second per detector. The afterglow peaks reach 1.02 in amplitude. The range $\tau <-150$\,nm is used to normalize the \gtau function to one.}
\label{fig.spurious_correlations}
\end{figure}

\begin{itemize}
    \item Time-to-digital converters (TDCs) generally suffer from an imperfect time-bin division, called ``differential nonlinearity'' (DNL): some bins are longer than others, in a repeatable way. This creates a spurious correlation pattern. This is usually repeatable and can thus be measured and subtracted from the data, as we did in \cite{Guerin:2017}. Alternatively, some TDC manufacturers use an internal calibration procedure to correct for this effect, at the cost of a slight decrease of the timing resolution. This is the case for the TDC we are using\footnote{The \emph{Time Tagger Ultra} from \emph{Swabian Instruments}.} since ref. \cite{Guerin:2018}. 
    \item TDCs also suffer from cross-talk between channels. For the TDC we are using, this creates spurious correlations on the order of $10^{-3}$ in amplitude (i.e., comparable to the bunching amplitude) for time lag between -100 and +100\,ns. To avoid that, we introduce a delay between the channels that we want to correlate, either an optical delay using fibers, or an electronic delay using a long coaxial cable between one of the SPADs and the TDC [Fig.\,\ref{fig.spurious_correlations}]. Another solution is to use two different synchronized TDCs at the different telescopes, record the arrival times of all photons and compute the correlations afterwards.
    \item Another spurious correlation is due to a phenomenon taking place in avalanche photodiodes, called ``afterglow'' or ``breakdown flash'' \cite{Kurtsiefer:2001} (not to be confused with afterpulsing, which does not introduce spurious correlations between two detectors, at least at first order): After a photon detection, the semiconductor can emit a flash of light. If this flash can propagate to the second detector, this will create a spurious correlation peak. This is the case, in particular, when the detectors are coupled to multimode fibers, and connected via a fibered splitter, due to small reflections at the interfaces after the splitter. Here also, this problem can be mitigated by using appropriate fiber lengths to ensure that the afterglow peaks are far enough from the bunching peak [Fig.\,\ref{fig.spurious_correlations}]. Note that this problem does not occur when the SPADs are not connected via a splitter, for example for correlations between two telescopes. It only occurs for the zero-baseline measurements.
    \item Finally, we have observed some spurious oscillatory correlations appearing as small oscillations centered around the bunching peak, i.e. at the optical zero delay, with a range of a few tens of ns. Their amplitude decreases when the SPADs are in RF-shielded boxes and put farther apart. We interpret this phenomenon as an electronic/radiative cross-talk: when one SPAD detects a photon, its circuit emits a radiation, which very slightly modifies the detection probability of the second SPAD. Typically, and with our shielding boxes, putting the SPADs $\sim 2-3$\,m apart is sufficient to make those oscillations not visible. But this can be a problem if the available space is limited (e.g. at a telescope).
\end{itemize}

In Fig.\,\ref{fig.spurious_correlations}, we show a scheme that puts the TDC cross-talk and the afterglow peaks far away from the bunching peak. We also show the corresponding \gtau measurements.

\subsubsection{SNR on an artificial star}

\begin{figure}[h]
\includegraphics{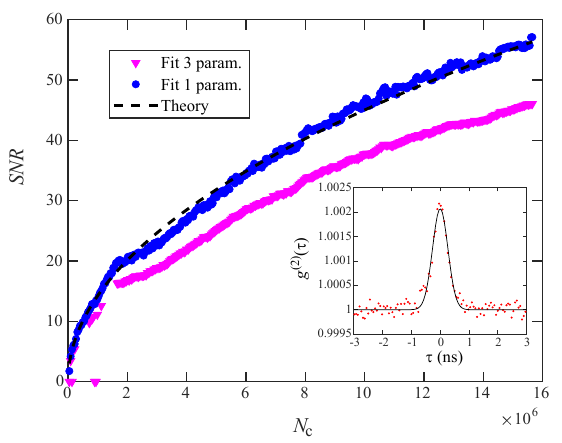}
\caption{Signal-to-noise ratio measured in the lab with an artificial star, as a function of the number of coincidences in the correlation histogram far from the bunching peak. The dashed line is the expected $S\!N\!R_\mathrm{fit} = \tau_\coh \sqrt{N_\mathrm{c}/(2\sqrt{\pi}\tau_\el t_\bin)}$ with $t_\bin = 10$\,ps. Inset: Bunching peak at the maximum integration time. The binning has been changed to $t_\bin = 60$\,ps for graphical improvement. The goodness of fit is $R^2 = 0.93$.}
\label{fig.SNR_exp}
\end{figure}

With the setup of Fig.\,\ref{fig.spurious_correlations}(a) we have realized a very long acquisition ($\sim 120$ h) to check the SNR formula and the possible deviation from it. The result is shown in Fig.\,\ref{fig.SNR_exp}: the agreement is excellent up to our maximum $SNR \sim 55$. Here, we plot the SNR as a function of the coincidence count in the correlation histogram (Eq. \ref{eq.Nc}).

\subsubsection{Practical considerations}\label{sec.practical}

Note that in all our measurements so far \cite{Guerin:2017,Guerin:2018,Rivet:2020,deAlmeida:2022,Matthews:2023}, we have used a three-parameter fit, which is not optimum. But how to fix the width and position of the bunching peak for the one-parameter fit?

Here, for Fig. \ref{fig.SNR_exp}, we have used the results of the three-parameter fit at the maximum integration time. This is clearly introducing a bias, as the statistical uncertainties on those parameters are neglected. Therefore, they produce systematic errors on the one-parameter fit.

The proper method would be to use a pulsed laser (with pulse duration much shorter than $\tau_\el$) to characterize the instrument response function with very high accuracy. We have done so, but we have observed, even in laboratory conditions, very small differences (on the order of 50 ps for both the position and width) between the bunching results and the pulsed-laser characterization. This is enough to make the one-parameter fit fail. It is known that the SPAD response depends on the wavelength and on the count rate, and we have taken care to have the same parameters for both acquisitions. However, it might also depend on the temperature or other less controlled parameters at the time scale of the long acquisition time. Thermal expansion in fibers and cables also influences the delays. 

These fluctuations are much worse in observing conditions at a telescope, in which the temperature can vary a lot, and the count rate is not stable due to changing turbulence conditions and atmospheric absorption. In addition, there is always some uncertainty in the computation of the optical path difference between the telescopes pointing at the same stars because of the uncertainty in the telescopes' positions, such that the bunching peak position is usually not known with picosecond precision.


Empirically, we observe, both using the numerical data of Fig. \ref{fig.SNR_num} and the experimental data of Fig. \ref{fig.SNR_exp}, a constant ratio
\begin{equation}
    \frac{S\!N\!R_\mathrm{3 param.}}{S\!N\!R_\mathrm{1 param.}} \approx 0.82 \,.
\end{equation}
The difference is not that large and is not worth taking the risk of introducing systematic errors by fixing two parameters. Therefore, in most practical cases, a fit with three parameters should be used and the useful equation for the SNR becomes
\begin{equation}\label{eq.SNR_practical}
S\!N\!R \approx 0.44  \frac{1}{2} \alpha N_\mathrm{ph}(\lambda) A |V(r)|^2 \sqrt{\frac{T_\mathrm{obs}}{\tau_\mathrm{el}}} = 0.44 \frac{|V(r)|^2}{2} \frac{F}{\Delta\nu} \sqrt{\frac{T_\mathrm{obs}}{\tau_\mathrm{el}}} \; ,
\end{equation}
where the numerical prefactor is $\approx 0.82/\sqrt{2\sqrt{\pi}}$ (we have left the $1/2$ to show that it is for unpolarized light; the $1/2$ disappears for polarized light). Note also that the factor $|V|^2/(2\Delta\nu)$ is simply the integral $I$ of the bunching peak. Then the equation
\begin{equation}\label{eq.SNR_final}
S\!N\!R \approx 0.44 \,I F \sqrt{\frac{T_\mathrm{obs}}{\tau_\mathrm{el}}} \; 
\end{equation}
is identical for polarized or unpolarized light, unresolved or partially resolved sources. $I$ and $\tau_\el$ are simply the integral and rms width of the fitted bunching peak. This is useful to compare to experimental data.

\subsubsection{On-sky validation of the SNR formula}

Let us compare the prediction from Eq.\,(\ref{eq.SNR_final}) with some of our previous published measurements, for instance those from \cite{Guerin:2018}. For the observation of Rigel, we had an observing time $T_\obs = 4.3$\,h, an average count rate $F = 1.8$\,Mcps, and the fitted bunching peak gave an rms width $\tau_\el = 350$\,ps and an area $\tau_\coh = 1.11 \pm0.2$\,ps, yielding an SNR of 5.6, while the application of Eq.\,(\ref{eq.SNR_final}) gives 5.8. For the observation of Vega, $T_\obs = 11.1$\,h, $F = 2.3$\,Mcps, $\tau_\el = 470$\,ps, and $\tau_\coh = 1.07 \pm0.11$\,ps, yielding an SNR of 9.7 against 10 predicted from Eq.\,(\ref{eq.SNR_final}). The measurements with stellar light are thus in very good agreement with the predictions.


\subsubsection{Systematic error}

The SNR discussed so far is only related to the statistical uncertainty of the fit, and thus does not include any systematic error. Even in the absence of any spurious correlation, there is one source of systematic error, which is that the bunching peak is not really Gaussian, but typically contains small but large tails, whose integral is not always negligible. These tails are usually hidden in the noise and thus not counted in the bunching peak area, which is therefore underestimated. The SNR formula based on the Gaussian response hypothesis is then inaccurate.

The exact instrument response function (IRF) can be experimentally characterized with a pulsed laser as input light. With our SPADs, there is a very clear deviation from a Gaussian. Fitting the IRF by a Gaussian leads to an underestimation of the integral by a few percent. However, even in our most precise bunching measurements (using a continuous source and integrating for several days), up to $S\!N\!R\sim 80$ \cite{Matthews:2022}, we have not observed any deviation from a Gaussian. We interpret this as being due to the small drifts during the acquisition, which makes the bunching peak more Gaussian than it should be.



Note that the absolute value of the bunching peak area is not really what matters for astronomy; it is instead its evolution as a function of the baseline. Therefore, this systematic error is mitigated by a careful calibration of the zero-baseline correlation that uses the same fitting procedure and the same couple of detectors as for the spatial correlation measurement.


\subsection{On-going developments: wavelength multiplexing with high-time resolution detectors}\label{sec.future}

In order to broaden the applications of SII beyond very bright stars, a significant enhancement of the current instruments' sensitivity is necessary. The most efficient is to use larger light collectors, but this is obviously not always possible. Minimizing losses or increasing the quantum efficiency of the detectors brings about a marginal improvement. Progress in the timing resolution of the detectors can yield a significant improvement but not orders of magnitude. In fact, it was already envisioned by Hanbury Brown that the only way to greatly improve the sensitivity of SII is to perform simultaneous measurements at many wavelengths \cite{HBT:1968}. This `multiplexing' approach has been further discussed in more recent papers (e.g., \cite{Trippe:2014,Lai:2018,Walter:2023}) but laboratory experiments along this line are only starting \cite{Tolila:2024,Ferrantini:2025}.

The idea is as follows: Since it is necessary to spectrally filter the light (see Sec.\,\ref{sec.SNR_HBT}), most of the stellar light is not used. Instead of throwing it away, one can use it for independent $g^{(2)}$ measurements at neighboring wavelengths. If the $N$ different measurements contain the same physical information, one can simply average them, and win a factor $\sqrt{N}$ on the SNR.
The possibility of using wavelength multiplexing with a large number of channels, yielding a significant increase in sensitivity, is highly dependent on the availability of high-time-resolution \emph{arrays} of detectors, such as SPAD arrays \cite{Milanese:2023} or SNSPD (superconducting nanowire single-photon detectors) arrays \cite{Chang:2019,Wollman:2024,Fleming:2025}, which have only recently been available, sparking the revival of the field.

The design of the spectral dispersion instrument 
is a delicate task and discussing it goes much beyond the scope of this article. Let us briefly mention that it has to have an excellent throughput, otherwise one already needs many channels only to break even. Moreover, one needs to be careful when using gratings because they add temporal dispersion \cite{Visco:2008}.
Furthermore, when computing the resulting expected SNR, one has to pay attention to remaining in the validity range of the equation. In particular, if the light is dispersed over a large number of spectral channels, the optical bandwidth may become very small for each channel, and consequently the coherence time becomes large. With timing resolution on the order of $\sim 20$\,ps or even faster \cite{Zadeh:2021,Gramuglia:2022}, the condition $\tau_\coh \ll \tau_\el$ may no longer be fulfilled. It is thus interesting to see how the SNR scaling evolves in that regime. Another critical point is that the spectra we correlate should be the same at the two (or more) telescopes, which becomes more challenging as the spectral channels get narrower. We examine these two points in the following.

\subsubsection{Beyond the hypothesis $\tau_\coh \ll \tau_\el$}

\begin{figure}[t]
\includegraphics{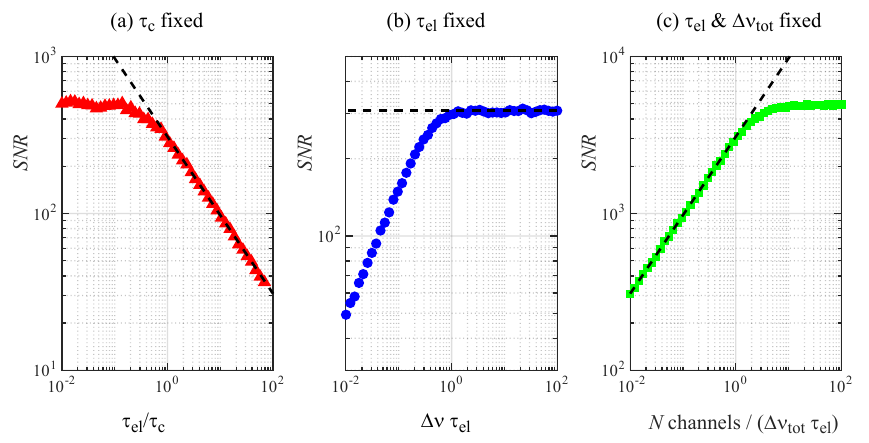}
\caption{Numerical simulations of the signal-to-noise ratio as a function of $\tau_\el$ and $\tau_\coh$ in different situations. (a) $\tau_\el$ is varied and $\tau_\coh$ is fixed. This corresponds to improving the resolution of the detector (towards the left). The SNR increases as $1/\sqrt{\tau_\el}$ as expected until it saturates when $\tau_\el \sim \tau_\coh$. The dashed line is $S\!N\!R = 0.44 \tau_\coh/(\sigma_\mathrm{noise}\sqrt{\tau_\el t_\mathrm{bin}})$, with $\tau_\coh = 1$, $t_\mathrm{bin} = 2\times10^{-3}$ and $\sigma_\mathrm{noise} = 10^{-2}$. (b) The detector performance is fixed ($\tau_\el = 1$) and we vary the optical bandwidth $\Delta\nu$. Although the $x$-axis is the same as in panel (a) in virtue of $\tau_\coh = 1/\Delta\nu$, the result is different because the number of detected photons increases with $\Delta \nu$. In the limit $\Delta\nu \, \tau_\el \gg 1$, we recover the well-known result that the SNR is independent of the optical bandwidth. The dashed line is the same equation as in (a) except that now $\sigma_\mathrm{noise} = 10^{-2}/\Delta\nu$. In the opposite regime, the SNR increases as $\sqrt{\Delta\nu}$. (c) We still consider $\tau_\el = 1$ and a fixed total wavelength range $\Delta\lambda_\mathrm{tot}$ or equivalently a total frequency range $\Delta\nu_\mathrm{tot}$, divided into $N$ independent channels. The computation is the same as in (b) but the SNR is multiplied by $\sqrt{N}$. However, above $N\sim \Delta\nu_\mathrm{tot}\,\tau_\el$, we enter the regime $\Delta\nu\,\tau_\el <1$ and the SNR saturates.}
\label{fig.SNR_future}
\end{figure}

We consider a Gaussian \gtau function\footnote{We take a Gaussian instead of a sinc$^2$ (Eq.\,\ref{eq.sinc}) to avoid inducing a systematic error in the fitting procedure. In a true experiment, the fit function would have to be adapted.} of amplitude one and area $\tau_\coh$, convolved with another Gaussian of area one and rms width $\tau_\el$. We add noise as previously and determine the SNR from a 3-parameter Gaussian fit. We report the results in Fig.\,\ref{fig.SNR_future} for three different situations. First, the coherence time is fixed, and we vary $\tau_\el$ [Fig.\,\ref{fig.SNR_future}(a)]. As expected, it shows that the SNR increases when the timing resolution is improved ($\tau_\el$ decreases), but the SNR saturates when $\tau_\el$ reaches $\tau_\coh$. Second, we fix $\tau_\el$, and we vary $\tau_\coh$ by varying the spectral bandwidth $\Delta \nu$ [Fig.\,\ref{fig.SNR_future}(b)]. In that case, the number of detected photons increases with $\Delta \nu$, but since $\tau_\coh$ decreases, in the limit $\tau_\coh \ll \tau_\el$, the SNR becomes independent of the bandwidth, as already discussed in Sec.\,\ref{sec.SNR_HBT}. However, in the opposite regime, the (lower) SNR increases as $\sqrt{\Delta \nu}$. This is because the photon flux increases linearly with $\Delta \nu$ and the width of the measured bunching peak is reduced linearly with $\Delta \nu$, as it is fully resolved. Since the fit uncertainty is roughly proportional to the square root of the number of points in the bunching peak, we finally obtain a $\sqrt{\Delta \nu}$ dependence in this regime. Finally, we can also fix the total wavelength range and divide it into $N$ spectral channels, and study how the SNR evolves with $N$ [Fig.\,\ref{fig.SNR_future}(c)]. We expect a $\sqrt{N}$ enhancement in the regime $\tau_\coh \ll \tau_\el$, but when $\tau_\coh > \tau_\el$, the SNR saturates. This happens when the number of channels is on the order of the ratio between the total optical bandwidth and the electronic bandwidth. Beyond that point, to continue improving the SNR, one has to increase the total optical bandwidth.

This limitation is implicitly taken into account in the proposal \cite{Walter:2023}, for example, since they consider individual bandwidths of $\Delta\lambda \approx 0.1$\,nm  at $\lambda \approx 500$\,nm with $N \approx 2000$ channels. This corresponds to $\tau_\coh \approx 8$\,ps, very close to the considered timing resolution $\tau_\el \approx 10$\,ps. But the total wavelength range is then $\Delta\lambda_\mathrm{tot} =$\,200\,nm, which surely makes the interpretation of the measured visibility a challenge.

Finally, note that there is also a fundamental limit in simultaneously determining the arrival time and frequency (via the spectral dispersion) of a photon \cite{Jirsa:2023}: $\Delta \nu \Delta t \geq 1/4\pi$ (Heisenberg or Fourier dispersion relation, where here $\Delta\nu$ and $\Delta t$ are standard deviations). This means that the passage through the dispersive optics induces a supplementary timing uncertainty. The order of magnitude is the same as the coherence time\footnote{ The exact numerical factors depend on the definitions and spectral shape. For instance, let us consider a Gaussian spectrum of rms width $\Delta\nu$ for simplicity. Then the rms width of the \gtau function is ($2\pi\sqrt{2}\Delta\nu)^{-1}$ \cite{Dussaux:2016} and the Fourier-limited timing uncertainty $\Delta t = (4\pi\Delta\nu)^{-1}$. The two differ by a factor $\sqrt{2}$.}. Therefore, it gives another reason to say that it is useless to improve the timing resolution of detectors below $\tau_\coh$.

\subsubsection{Effect of mismatched spectra}

A common assumption is that the detected spectral flux, for a given $g^{(2)}$ measurement, is identical at all telescopes. In practice, there will be slight variations due to the instrumental setup. Left unaccounted, these variations will appear as a systematic error in the squared visibility measurement, which will be lower than with perfectly matched spectra. Moreover, it also affects the SNR, since 
all detected photons contribute to the noise but only the overlapping spectral bands are correlated. In the computation of the SNR, the bunching peak amplitude (Eq.\,\ref{eq.ampl}) is proportional to the coherence time $\tau_\coh$ (Eq.\,\ref{eq.tau_c}). In case of mismatched spectra, $\tau_\coh$ should be replaced by a ``mutual coherence time''
\begin{equation}\label{eq.tau_mc}
\tau_\mathrm{mc} = \int |s_1(\nu)\,s_2(\nu)| d\nu \,,
\end{equation}
where $s_1(\nu), s_2(\nu)$ are the two spectra.

For example, let us consider two flat spectra of width $\Delta\nu$ with a small shift $\delta\nu$. The width of the overlapping band is $\Delta\nu-\delta\nu$ and the  mutual coherence time is
\begin{equation}\label{eq.tau_mc}
\tau_\mathrm{mc} = \frac{1}{\Delta\nu^2} \left(\Delta\nu-\delta\nu\right) = \tau_\coh \left(1-\frac{\delta\nu}{\Delta\nu}\right).
\end{equation}
Consequently, the SNR is reduced by a factor $1-\delta\nu/\Delta\nu$.


Note that this is equivalent to having background light. Indeed, let us consider the overlapping spectral band as the star light (giving the usual SNR, since the SNR is independent of the spectral bandwidth) and the nonoverlapping bands as uncorrelated background light. The effect of the sky background has been studied by HBT \cite{HBT:1958} and others \cite{Rou:2013} and they have shown that the SNR is proportional to the ratio star/(star+sky), which gives in our case $(\Delta\nu-\delta\nu)/\Delta\nu$, yielding the same result.

\section{Summary}

We have presented, in Sec.\,1, a short review of stellar intensity interferometry, wrapping up the physical principles, historical development, and current state of the art. It is a powerful technique that is undergoing a strong revival, thanks to the progress in detector technologies.

Yet, the main drawback of intensity interferometry is its poor sensitivity, a problem that can only be mitigated by using big collectors, long integration time, and a carefully optimized setup. To this end, it is absolutely necessary to use the best available detectors, minimize losses, and prevent spurious correlations. Therefore, it is helpful to know exactly what the SNR should be, depending on the experimental parameters, and to understand all the possible problems (rarely discussed in publications) that prevent reaching the theoretically expected SNR. Although HBT did a tremendous job in explaining in great detail how their interferometer worked \cite{HBT:book}, we felt that an update and adaptation to the photon-counting regime would be useful to the community. This is what we have provided in the second, quite technical, part of this article. Furthermore, we have discussed the on-going efforts to significantly enhance the sensitivity, including some fundamental limits. Not only it should be useful to experimentalists working in this field, but it also puts on firmer ground the extrapolations or proposals that one may make about using intensity interferometry to study faint objects \cite{Walter:2023,Dalal:2024,Kim:2025}.

We would like to thank Farrokh Vakili for his seminal inputs and guidance throughout the first, almost ten years of our work. We also thank Paul Stankus for his thorough review of the manuscript.



\def\bysame{\leavevmode ---------\thinspace}
\makeatletter\if@francais\providecommand{\og}{<<~}\providecommand{\fg}{~>>}
\else\gdef\og{``}\gdef\fg{''}\fi\makeatother
\def\cdrandname{\&}
\providecommand\cdrnumero{no.~}
\providecommand{\cdredsname}{eds.}
\providecommand{\cdredname}{ed.}
\providecommand{\cdrchapname}{chap.}
\providecommand{\cdrmastersthesisname}{Memoir}
\providecommand{\cdrphdthesisname}{PhD Thesis}

\end{document}